\documentclass[runningheads]{llncs}
\usepackage[ruled,vlined]{algorithm2e}
\usepackage{float}
\usepackage{graphicx}
\usepackage{amsfonts}
\usepackage{amsmath,amssymb}
\begin{document}
\title{Model-based Pre-clinical Trials for Medical Devices Using Statistical Model Checking}
%
%
\author{Haochen Yang\inst{1}\orcidID{0000-0001-9145-8575} \and
Jicheng Gu\inst{1}\orcidID{0000-0003-2982-1964} \and
Zhihao Jiang\inst{1,2}\orcidID{0000-0002-6730-6915}}
\authorrunning{H. Yang et al.}
%
\institute{ShanghaiTech University, Shanghai, China \and
Shanghai Engineering Research Center of Intelligent Vision and Imaging
\email{\{yanghch,gujch,jiangzhh\}@shanghaitech.edu.cn}}
\maketitle              
\begin{abstract}
Clinical trials are considered as the golden standard for medical device validation. 
However, many sacrifices have to be made during the design and conduction of the trials due to cost considerations and partial information, which may compromise the significance of the trial results.
In this paper, we proposed a model-based pre-clinical trial framework using statistical model checking.
Physiological models represent disease mechanism, which enable automated adjudication of simulation results.
Sampling of the patient parameters and hypothesis testing are performed by statistical model checker.
The framework enables a broader range of hypothesis to be tested with guaranteed statistical significance, which are useful complements to the clinical trials.
We demonstrated our framework with a pre-clinical trial on implantable cardioverter defibrillators.

\end{abstract}
\section{Introduction}
With the development of science and technologies, modern medical devices become more "intelligent", such that critical diagnosis and therapeutic decisions can be made by the device algorithm without interventions from medical professionals.
These \textit{Medical Cyber-Physical Systems (MCPS)} \cite{MCPS,MCPS_challenges} usually aim to deliver timely therapies during high risk or life-threatening conditions, so that patients with the devices can have improved life quality without the need to stay in hospital.
For instance, \textit{Implantable Cardio-verter Defibrillator (ICD)} detects irregular heart rhythms, and deliver therapies like pacing and defibrillation to restore normal heart rhythm \cite{ICD}.
\textit{Responsive NeuroStimulation (RNS)} systems detect abnormal brain activities and deliver high-frequency electrical stimulation to terminate seizures or Parkinson syndromes \cite{RNS}.

The medical device industry is highly regulated. 
Due to the autonomy and high risk therapies, MCPSs are classified as Class III devices, and the safety and efficacy of MCPSs have to be rigorously validated in clinical settings before the devices can enter the market \cite{WHO}.
Due to the large variability of patient conditions, diagnosis in these MCPSs cannot be 100\% accurate.
Therapies may be absent during high-risk conditions (false-negative), and therapies may be delivered during normal or low-risk conditions (false-positive).
Device manufacturers and regulators need clinical evidence that MCPSs can reduce the number of false-positives and false-negatives as low as possible \cite{Med_Dev}.

Cost is a primary consideration when designing clinical trials.
A clinical trial takes many years to complete and costs millions of dollars \cite{Cost_CT}.
Therefore the experiments in these clinical trials are carefully designed to achieve statistical significance with the least amount of enrolled patients.
In order to strike a balance between statistical significance and the size of enrolled patients, assumptions are made regarding the device under evaluation and the target population.
For example, the new device is X\% better in terms of safety/efficacy compared to the standard care.
These assumptions are made based on prior knowledge from previous clinical studies, animal tests or clinical literature, which may not reflect the ground truth.
Sacrifices are also made to the endpoint of the clinical trials, so that the duration of the trials and the number of patients enrolled are financially feasible. 
For example, instead of choosing "survival rate of patients" as endpoint of a clinical trial, the trial designer may choose "first time for inappropriate therapy", which has much shorter expected trial duration.
As the result, although clinical trials are the golden standard for clinical evidence, their results may not be directly useful to guide clinical decisions due to the trade-offs.
If wrong assumptions are made during the design of clinical trials, the trial may fail to generate result with statistical significance, essentially wasting all the money and time invested \cite{Trails_F}.
\begin{figure}[t]      
\centering      
\includegraphics[width=0.9\textwidth]{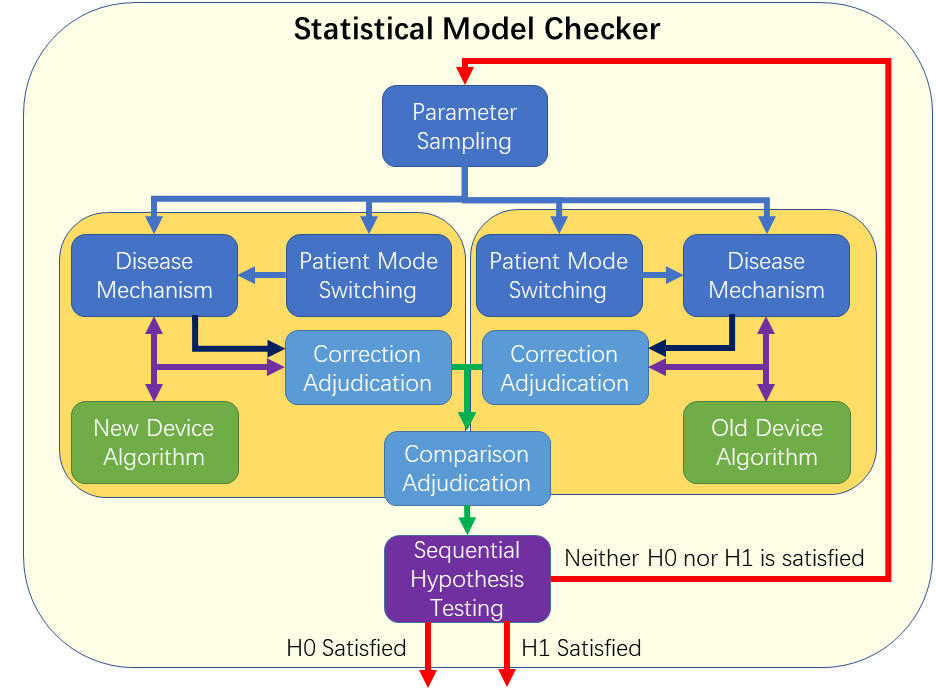} 
\caption{Overview of Our Proposed Framework}  
\label{Fig:overview}
\end{figure}

Physiological models have been proposed to provide safety and efficacy evidence for medical devices.
Closed-loop model checking of medical device software can provide formal and rigorous evidence, but may not be scalable to more complex algorithms \cite{STTT13}.
Model-based (in-silico) pre-clinical trials use physiological models as virtual patient, and have been used to complement and support clinical trials \cite{embc16,AP}.
However, these pre-clinical trials do not fully align with the statistical approaches in clinical trials, and may not be able to generate clinically-relevant results that can guide clinical decisions.


In this paper, we propose a pre-clinical trial framework for medical devices using statistical model checking.
The framework is illustrated in Fig. \ref{Fig:overview}.
A \textit{Disease Mechanism} module captures physiology of disease mechanism which can 1) interact with device algorithm in closed-loop, and 2) provide physiological context for adjudication.
A \textit{Patient Mode Switching} module captures the switching between patient conditions by changing the parameters of the physiological model.
A \textit{Parameter Sampling} module randomly samples a set of physiological parameters for the physiological model as well as the patient mode switching module.
Two virtual patients with the same set of parameters are used to evaluate two different devices for better comparability.
A \textit{Correctness Adjudication} module provides quantified evaluation of a simulation.
A \textit{Comparison Adjudication} module compares the quantified evaluation of the two device algorithm, and the \textit{sequential hypothesis testing} block inside statistical model checker UPPAAL SMC \cite{UPPAALSMC} then determines whether the hypothesis has been accepted with statistical significance after each iteration. 
The framework can be used to test hypothesises regarding the difference between a proposed device algorithm and an older algorithm with guaranteed statistical significance.
These hypothesises include not only the ones tested in clinical trials, but also the ones interesting but not feasible in clinical trials.
The result of these hypothesis testings can provide guidance to future clinical trials, including predicting trial results as well as validating assumptions during clinical trials.

The contribution of this paper is 3-folds: 1) We proposed a framework of pre-clinical trials for MCPSs using statistical model checking; 2) we proposed a set of clinically-relevant hypothesises that can be tested with our proposed framework; 3) we performed a series of pre-clinical trials on two algorithms for implantable cardioverter-defibrillator. 

The rest of this paper is arranged as follows: in Section 2, the statistical approaches used in clinical trials are introduced, and the limitations of clinical trials are demonstrated using an example trial; in Section 3, our proposed pre-clinical trial framework using statistical model checking is introduced; in Section 4, we demonstrate our proposed framework using a case study; we then end the paper with discussion and future work.

\section{Clinical Trials for MCPSs}
In this section, we introduce the statistical approaches used in clinical trials and use an example to illustrate their limitations. 

\subsection{Clinical Trial Basics}
There are several key properties of a clinical trial, which can affect its cost and clinical significance.
These properties are determined before the conduction of the trial.
\subsubsection{Endpoint}
An \textit{endpoint} of a clinical trial is a well-accepted, easily-observable and quantifiable criterion that reflects the purpose of the trial \cite{FDA_endpoint,Endpoint}.
For example, if a device is designed to treat a life-threatening disease, the death of patients due to the disease would be an appropriate endpoint for the clinical trial.
The endpoint affects the duration of the trial which is closely related to the cost of the trial.
For instance, if the death of patients is the endpoint of a trial and the patients in the target population are relatively young and healthy, the duration of the trial may be too long to be feasible.
Therefore sacrifices have to be made when choosing the endpoint in order to ensure reasonable trial duration.  
\subsubsection{Randomized Control Trials}
The objective of a clinical trial is to determine whether the device under evaluation is better in certain aspects compared to the standard care device.
Since it is infeasible to evaluate two devices on the same patient, the evaluation has to be performed on different patients, which are not directly comparable. 
Therefore the recruited patients are assigned into two groups: the treatment group with the new device, and the control group with the standard care device.
In order to make the two groups comparable, patients are randomly assigned into the groups to avoid bias \cite{Random}. 
\subsubsection{Hypothesis Testing in Clinical Trials}
In clinical trials, \textit{hypothesis testing} is used as the primary statistical method, and \textit{superiority hypothesis testing} is usually used for medical device trials \cite{SampleSize_book}.
The null hypothesis $H_0$ for superiority trial claims that the new device and standard care device have no difference, and the corresponding alternative hypothesis $H_1$ claims that the treatment group is superior than the control group in terms of the trial endpoint. 
Hypothesis testing aims to prove the correctness of the alternative hypothesis by rejecting the null hypothesis.
The significance level of the test $\alpha$ is first determined such that:
$$\alpha = \text{max } \mathbb{P}(\text{reject } H_0|H_0 \text{ is true})$$
Based on the test statistics from the significance test, the value and distribution of the probability for $H_0$ to be accepted (the p value) can be determined.
If the p value is larger than $\alpha$, there is not enough evidence to reject $H_0$; and if the p value is less or equal to $\alpha$, $H_1$ is accepted.
Traditionally $\alpha$ is set to 0.05 in hypothesis testings.
    
    
    



\subsubsection{Sample Size for Clinical Trials}
In clinical trials, the number of patients required to guarantee statistical significant results (the sample size) is determined before the trial.
For superiority trials, the desired sample size is related to how much better the new device is compared to the old device.
Since this information is not available before the trial, trial designers have to make an assumption (the new device is better than the old device by $d$) base on prior knowledge like literature review. 
$d$ is referred to as the \textit{effect size}, which can significantly affect the sample size estimation \cite{SampleSize_book}.
For example, reducing the effect size by half will increase the sample size by 4 times \cite{ss_effect}.
Sample size is also related to the significance level and power of the trial, as well as the estimated drop-off rate of the patients.
The sample size is the most important parameter of a clinical trial, as it directly determines the cost of the trial and the statistical significance of the trial result.

    

\subsection{The RhythmID Head-to-Head Trial (RIGHT)}
We use a previous clinical trial on implantable cardioverter defibrillator to illustrate the limitations of clinical trials.
\subsubsection{Implantable Cardioverter-Defibrillators}
Coordinated contractions of the heart is governed by electrical signals.
Anomalies in generation and conduction of electrical signals can cause irregular heart rhythm, which is referred to as arrhythmia.
The severity of arrhythmia depends on the locations of the anomalies.
Fast ventricular rate is referred to as ventricular tachycardia (VT), which can be fatal.
On the other hand, fast atrial rate is referred to as Supra-Ventricular Tachycardia (SVT), which is non-fatal.
Implantable Cardioverter-Defibrillators (ICD) are designed to detect and treat VT with anti-tachycardia pacing or defibrillation \cite{ICD}.
Treatments during SVT are considered as inappropriate therapies.
Modern ICDs are equipped with SVT-VT discrimination algorithms to reduce the number of inappropriate therapies while ensuring all VTs are treated.
\subsubsection{Background}
In 2009, the ICD manufacturer Guidant has developed a new SVT-VT discrimination algorithm called RhythmID.
Based on estimations before the clinical trial, RhythmID should have less risk of inappropriate therapies compared to PRLogic, which is an algorithm developed by Medtronic.
Therefore Guidant proposed the RhythmID Head-to-Head Trial (RIGHT) \cite{RIGHT} to validate this hypothesis, which is a randomized, unblinded, two-arm, prospective clinical trial.
\subsubsection{Endpoint}
The primary endpoint of RIGHT is the time of the first inappropriate therapy.
It can be categorized as a "Event-free Survival" endpoint, which generally results in relatively shorter trial duration and smaller sample size.
However, patients receiving inappropriate therapy once are considered as dead when using this endpoint, which is not the case for inappropriate therapies of ICD and may compromise the clinically-relevance of the result. 
"The rate/chance of inappropriate therapy over X years" is a clinically more interesting endpoint, but maybe infeasible for a clinical trial due to longer trial duration and larger sample size. 

\subsubsection{Sample Size}
1962 patients were enrolled in RIGHT.
Based on device manufacturer (Guidant (GDT) vs. Medtronics (MDT)) and type of device (Single chamber vs. Dual chamber), patients enrolled in RIGHT are divided into 4 separate groups: 507 for GDT Single, 478 for GDT Dual, 503 for MDT Single and 474 for MDT Dual.
The sample size was calculated based on the assumption that "the GDT algorithm has 25\% less risk for inappropriate therapies compared to the MDT algorithm".
The relatively large assumed difference results in smaller sample size, but may risk the statistical significance if the assumption is wrong.

\subsubsection{Hypothesis Testing of RIGHT}
Since the endpoint for RIGHT is related to how long a patient can survive (with no inappropriate therapies), the hypothesis testing is based on \textit{survival analysis}.
The survival function $S(t)$ is a function of the number of patients who received inappropriate therapy before time t.
In RIGHT, the alternative hypothesis is  $\mathrm{H}_{\mathrm{a}}:\mathrm{S}(\mathrm{t})_{\mathrm{GDT}} \neq \mathrm{S}(\mathrm{t})_{\mathrm{MDT}}$.
\subsubsection{Adjudication of Recorded Traces}
The saved execution traces of the ICDs are analyzed by three physicians to determine whether and when an inappropriate therapy has happened.
An adjudication is made if two out of three physicians have an agreement, otherwise the trace is discarded.
The lack of physiological context in the execution traces affects the accuracy of adjudication, which can further affect the trial results.

\subsubsection{Result}
At the end of RIGHT, results showed that the MDT group has longer survival time compared to the GDT group.
The GDT device has 34\% higher risk for inappropriate therapies compared to the MDT algorithm (P=0.003),which is the opposite to the initial assumption \cite{RIGHT_result}.

\subsection{Limitations of Clinical Trials}
From the example above, the limitations of clinical trials can be summarized as follows:
\subsubsection{Cost-sensitive:}
Clinical trials are very costly and many sacrifices have to be made in order for a trial to be feasible.
The endpoint sacrifices physiological-relevance for trial duration.
Bold superiority assumptions are made in order to maintain reduced but feasible sample size.
All these sacrifices may compromise the information we learn from the trial.
\subsubsection{Sampling from Target Population:} 
In clinical trials, the patients recruited may not be a good representation of the target population.
Patients with rarer conditions may not be recruited, and will not yield statistical significant result due to the rarity.
Sampling is usually performed on meta-data like gender and age, instead of physiological parameters, which may affect the comparability between groups.

\subsubsection{Comparison:}
Since it's not feasible to evaluate two devices on the same patient at the same time, the comparisons between devices are made indirectly using statistical methods.
\subsubsection{Adjudication:}
Adjudications of clinical trials on MCPSs are made by the physicians on the recorded execution traces of the devices.
Due to the embedded nature of most MCPSs, only segments of device executions can be recorded.
The recorded traces also lack physiological context, which may affect the adjudication results.

\section{Pre-clinical Trials with Statistical Model Checking}
The limitations of clinical trials can be complemented with model-based approaches.
In this section, we introduce our proposed model-based framework for pre-clinical trials using statistical model checking.

The overview of the framework is illustrated in Fig. \ref{Fig:overview}.
Physiological model of the patient provides physiological context which enables automated and accurate adjudication.
The cost of physiological model simulation is negligible compared to the clinical trials.
Target population can be sampled base on physiological parameters of the physiological model, which can better cover rare patient conditions.
Moreover, devices can be evaluated on the same "patient", or "patient" with the same physiological parameters.
Finally, statistical significance of the results can be guaranteed using sequential statistical model checking. 
\subsection{Patient Model}
The physiological model of the patient can be separated into two modules.
The disease mechanism module captures physiological mechanism of the disease to interact with the device algorithm.
The mode switching module captures the different conditions or modes of the patient which correspond to correct/incorrect diagnosis and therapy for easy adjudication.
The parameters of the model should reflect physiological variability of the patients.
\subsection{Parameter Sampling}
The parameters of the patient model should be given as distributions, and statistical model checker samples the parameters independently to create a virtual patient in each iteration.
Two virtual patients with the same parameters evaluate both the new device and the standard care device to provide better comparability than in clinical trials.

\subsection{Adjudication}
The adjudication module has access to the state of the patient model, so that quantitative and qualitative assessment of the execution traces can be performed.
The adjudication module contains two parts.
The correction adjudication evaluates each execution trace from one device, and comparison adjudication compares the evaluations of execution traces from the two devices.
The result of the adjudication is used by the statistical model checker to determine whether the trial has achieved desired statistical significance.

\subsection{Statistical Model Checking}
Statistical Model Checker performs hypothesis testing on stochastic systems. 
The key idea is to infer whether the system satisfies the quantitative properties by monitoring simulations of system, and use hypothesis testing to provide statistical evidence.

\begin{figure}      
\centering      
\includegraphics[width=0.35\textwidth]{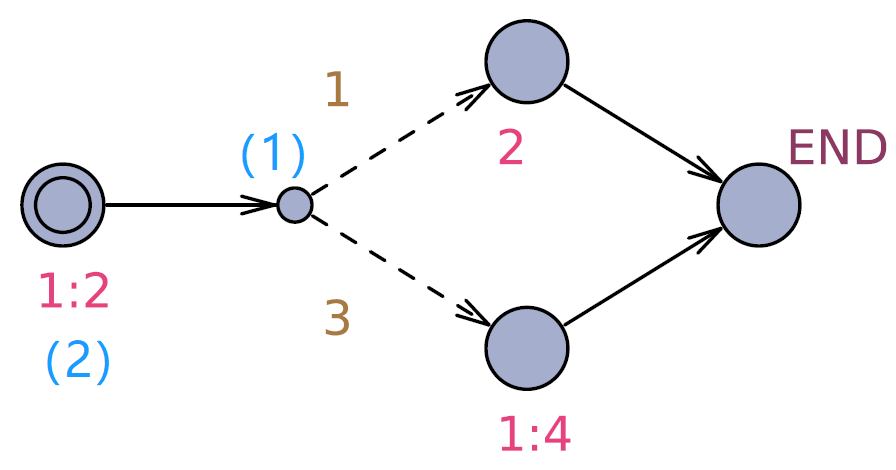}     

\caption{An example of timed automaton in UPPAAL SMC}  
\label{Fig:SMC}
\end{figure}

In this project, we use the UPPAAL SMC \cite{UPPAALSMC} to implement our framework. 
UPPAAL SMC use an extension of the Timed Automata (TA) \cite{timed_automata} formalism.
Stochastic behaviors can be modelled using probabilistic transitions (Fig. \ref{Fig:SMC}.(1)), as well as rate of exponential on locations (Fig. \ref{Fig:SMC}.(2)). 

UPPAAL SMC can perform statistical model checking using several queries.
In this paper, we primarily use the following query:
$$Pr[ bound1 ]( psi1 ) >= Pr[ bound2 ]( psi2 )$$
which compares the probability for given system $\mathcal{A}$ to satisfy the property $\varphi_1$ within bound $bound1$, with the probability for given system $\mathcal{A}$ to satisfy the property $\varphi_2$ within bound $bound2$. 




For given two properties $\varphi_1,\varphi_2$. The null and alternate hypothesis are shown as following.
$$ H_0:p_1=\mathbb{P}_{\mathcal{A}}(\diamond_{t_1\leq T_1}\ \varphi_1) \geq p_2 = \mathbb{P}_{\mathcal{A}}(\diamond_{t_2\leq T_2}\ \varphi_2)$$ 
$$ H_1:p_1=\mathbb{P}_{\mathcal{A}}(\diamond_{t_1\leq T_1}\ \varphi_1) < p_2 = \mathbb{P}_{\mathcal{A}}(\diamond_{t_2\leq T_2}\ \varphi_2)$$ 
Probability comparison can be reduced to testing the hypothesis $H_0$ against $H_1$. 

In UPPAAL SMC, Extend Wald's Sequential Hypothesis Testing \cite{ExtendWald} is used to test the hypothesis. 
Sequential Hypothesis Tesing is an iterative process.
In each iteration, the system $\mathcal{A}$ executes for two independent simulations $r_1$ and $r_2$, and $\varphi_1, \varphi_2$ are evaluated in their corresponding runs. 
The result for this iteration is a pair $(x_1,x_2)$, in which $x_1=1$ if $\varphi_1$ is satisfied in $r_1$, and $x_1=0$ otherwise.
The cumulative result is then analyzed by SMC.
The system enters another iteration if neither $H_0$ nor $H_1$ is accepted, and terminates when either $H_0$ or $H_1$ is accepted.

Compared to the hypothesis testing approach used in clinical trials, the sequential hypothesis testing used in UPPAAL SMC does not need to decide the sample size beforehand, and can provide guaranteed statistical significance.
However, it is usually infeasible to use sequential statistical testing in clinical trials for medical devices, in which the duration for each patient's enrollment usually takes many years. 


\section{Case Study: Pre-clicnial Trials for RIGHT Using Statistical Model Checking}

\subsection{Physiological Model}
\begin{figure}      
\centering      
\includegraphics[width=0.4\textwidth]{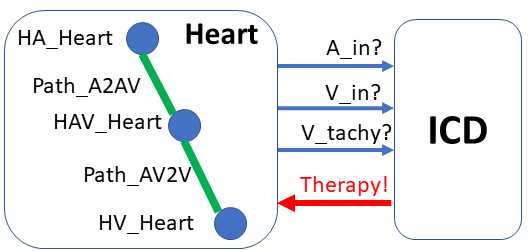}     
\caption{Interface between the heart and the ICD}  
\label{Fig:interface}
\end{figure}

A dual chamber ICD receive three channels of electrical signals from the heart.
Through the sensing circuit and algorithm, the \textit{Atrial channel} and the \textit{Ventricular channel} report whether there are atrial and ventricular contractions within the heart, which correspond to the $A\_in$ and $V\_in$ events.
The \textit{Shock channel} senses far-field ventricular electrogram, and through the morphology comparison algorithm, the ICD can judge whether a $V\_in$ event correspond to a VT event ($V\_tachy$).
The $Therapy$ event from the ICD resets the states of the heart model, mimicking the effect of defibrillation. 
The interface between the heart model and the ICD algorithm is shown in Fig. \ref{Fig:interface}.

\begin{figure}      
\centering      
\includegraphics[width=0.7\textwidth]{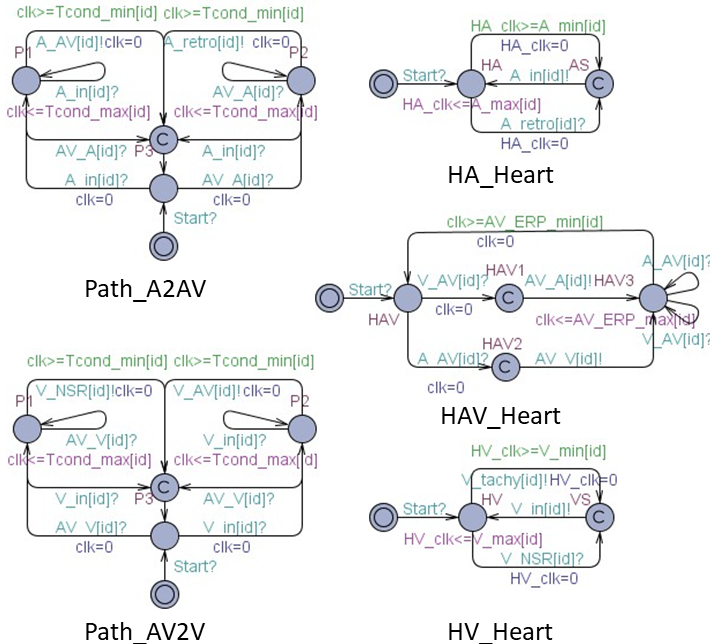}     
\caption{Timed automata heart model}  
\label{Fig:heartmodel}
\end{figure}
In this case study, we have adopted the Virtual Heart Model (VHM) \cite{VHM} to model the interaction between the heart and the ICD, as well as disease mechanism.
The VHM is a timed-automata-based heart model which is capable of simulating the electrical activities of various heart conditions, and has been used to model the electrical activities of the heart during model checking and closed-loop testing of implantable cardiac devices \cite{STTT13,embc16}.
The VHM model used in this case study is illustrated in Fig. \ref{Fig:heartmodel}.
The two node automata $HA\_Heart$ and $HV\_Heart$ model the generation of the three input events to the ICDs.
The two path automata $Path\_A2AV$ and $Path\_AV2V$ model the conduction between node automata.
The node automaton $HAV\_Heart$ models the blocking property of the AV node.
The heart model is capable of illustrating the origin and conduction pattern of electrical signals within the heart, which provides enough physiological context for the adjudication.
    
    
\begin{figure}      
\centering      
\includegraphics[width=0.9\textwidth]{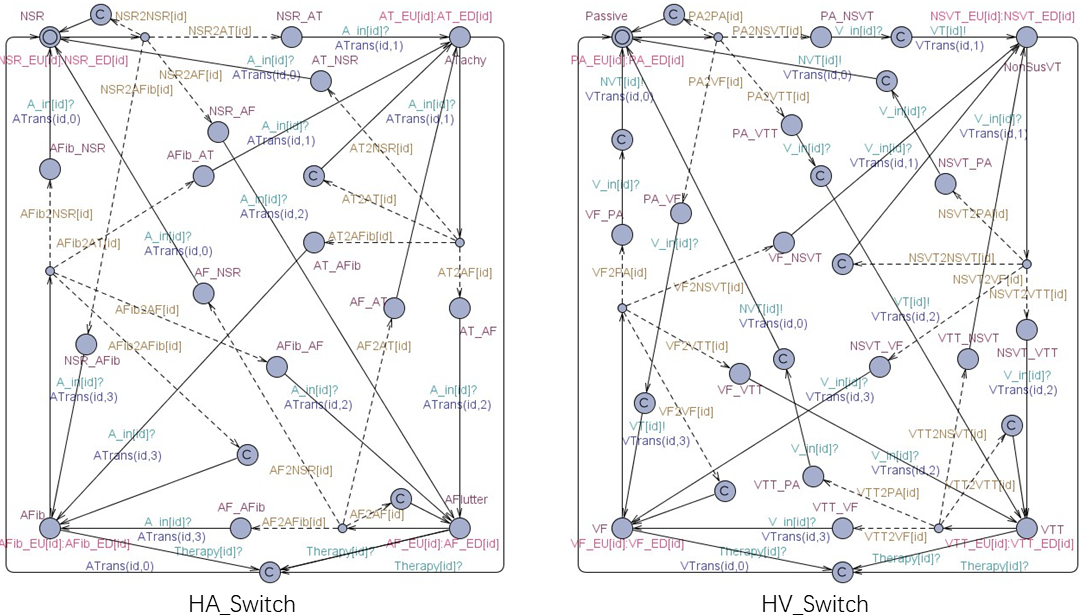}    

\caption{Mode Switching model for atria and ventricles} \label{Fig:switch} 
\end{figure}
    
\subsection{Patient Mode Switching Model}
The switching between tachycardia and normal heart rhythm is relatively quick, and tachycardia can occur in either atria or ventricles independently. 
Therefore we created two mode switch models $HA\_Switch$ and $HV\_Switch$ to switch the parameters of the heart model, so that combinations of atrial and ventricular tachycardia can be modeled (Fig. \ref{Fig:switch}).
Each mode has a duration, which is governed by the rate of exponential of the locations.
Switching among modes are governed by probabilistic transitions.

    

\subsection{Parameter Sampling}
The mode switching model and the physiological model have 68 parameters in total, and are independently sampled from normal distributions.
The distributions of parameters were obtained from literature review \cite{josephson}.
Since the patients are in Normal Sinus Rhythm (NSR) 99\% of the time, we shortened the duration of the NSR state which can significantly increase the effective duration of the trial. 

\subsection{Adjudication}
There are two adjudication modules in this case study:
The correctness adjudication module determines whether a therapy is appropriate.
The comparison adjudication module determines which algorithm is better in various criteria, such that hypothesis testing can be performed.
\begin{figure}      
\centering      
\includegraphics[width=0.9\textwidth]{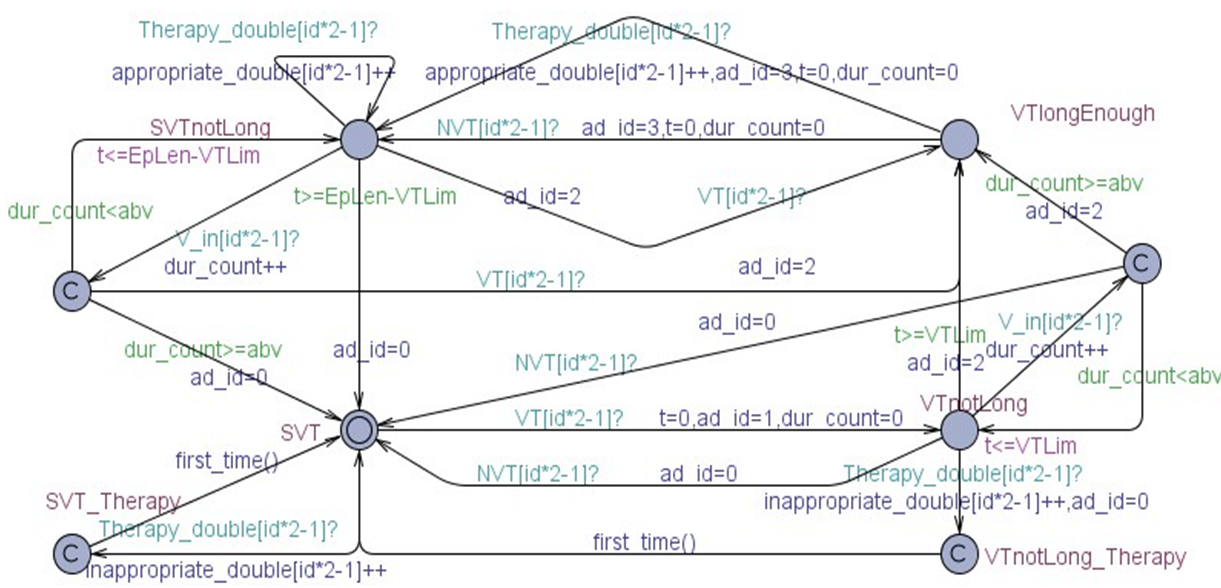}     

\caption{Correction adjudication module determines 
whether a therapy is appropriate}  
\label{Fig:adjudication}
\end{figure}

The correctness adjudication module for this study is shown in Fig. \ref{Fig:adjudication}.
It utilizes the state of the patient mode switch model to determine whether there has been a persistent ventricular tachycardia, which cannot be obtained just from the execution traces recorded by the devices.
The number of inappropriate therapies and the time for first inappropriate therapy are saved for hypothesis testing.

\begin{figure}      
\centering      
\includegraphics[width=0.5\textwidth]{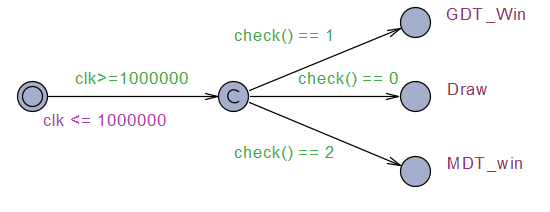}     

\caption{Comparison adjudication module compares the evaluation of the two algorithms}  
\label{Fig:comparison}
\end{figure}
The comparison adjudication module is shown in Fig. \ref{Fig:comparison}.
After the simulation finishes, a function $check()$ evaluates whether there exists difference between the two algorithms.
Then the following query was checked:
$$Pr[bound] (<>GDT\_Win) \geq Pr[bound] (<> MDT\_Win)$$

\subsection{The VT-SVT Discrimination Algorithms}    
In this case study, we use the RhythmID algorithm from GDT and PRLogic algorithm from MDT presented in \cite{embc16}, which were inferred from algorithm descriptions in open literature.
The algorithms are converted from Matlab code to timed automata \cite{timed_automata}, which are attached in the Appendix for interested readers.
Since saving clock values is not allowed in timed automata, we had to sacrifice the symbolic state representation within timed automata in order to model the algorithm.
As the result, traditional model checking on these algorithms is not feasible due to state explosion problems.

\subsection{Pre-clinical Trials for RIGHT}
A series of pre-clinical trials are conducted using the framework we proposed.
All trials are conducted on the dual-chamber version of the ICD algorithms.
\subsubsection{Trial 1: Compare Chance of Inappropriate Therapy}
The chance of inappropriate therapy on any patient within a prolonged period of time is a clinically-relevant criterion to compare two algorithms, but is not feasible in clinical trials as two devices cannot be evaluated on the same patient.
The hypothesis under evaluation are: 
    $$H_0:\mathbb{P}(\diamond t\leq T\ N_{MDT}>0) \geq \mathbb{P}(\diamond t\leq T\  N_{GDT}>0)$$
    $$H_1:\mathbb{P}(\diamond t\leq T\ N_{MDT}>0) < \mathbb{P}(\diamond t\leq T\  N_{GDT}>0)$$
In which $T=1000000$ms and $N_{GDT}$ and $N_{MDT}$ represent the number of inappropriate therapies when the same patient has GDT and MDT devices, respectively.
In UPPAAL SMC, the comparison adjudication module is programmed such that:
    \[ check() =
  \begin{cases}
    1       & \quad \text{if } N_{GDT}=0 \& N_{MDT}>0\\
    0  & \quad \text{if } (N_{GDT}=0 \& N_{MDT}=0) || (N_{GDT}>0 \& N_{MDT}>0)\\
    2  & \quad \text{if } N_{GDT}>0 \& N_{MDT}=0
  \end{cases}
\]
After 582 simulations, $H_0$ was rejected and $H_1$ was accepted with 95\% confidence and 95\% power, suggesting that the new algorithm from GDT has a higher chance of inappropriate therapy compared to the MDT algorithm, which aligns with the result of RIGHT.
    
\subsubsection{Trial 2: Event-free Survival Time}
Event-free survival time is a common endpoint for clinical trials.
The survival time for a patient $ST$ is defined as:
\[ ST =
  \begin{cases}
    T_{sim}       & \quad \text{if no inappropriate therapy occurred during simulation} \\
    T_{inapp}  & \quad \text{if inappropriate therapy occurred during simulation}
  \end{cases}
\]
in which $T_{sim}$ is the total simulation time, and $T_{inapp}$ is the time of the first inappropriate therapy.
We compare whether GDT has better survival time compared to MDT with the following hypothesis:
    $$H_0:\mathbb{P}(\diamond t\leq T\ ST_{GDT}>ST_{MDT}) \geq \mathbb{P}(\diamond t\leq T\  ST_{GDT}<ST_{MDT})$$
    $$H_1:\mathbb{P}(\diamond t\leq T\ ST_{GDT}>ST_{MDT}) < \mathbb{P}(\diamond t\leq T\  ST_{GDT}<ST_{MDT})$$
The comparison adjudication module is programmed such that:
    \[ check() =
  \begin{cases}
    1       & \quad \text{if } ST_{GDT}>ST_{MDT}\\
    0  & \quad \text{if } ST_{GDT}=ST_{MDT}\\
    2  & \quad \text{if } ST_{GDT}<ST_{MDT}
  \end{cases}
\]   
After 582 simulations, $H_0$ was rejected and $H_1$ was accepted with 95\% confidence and 95\% power, suggesting that the new algorithm from GDT has shorter event-free survival time compared to the MDT algorithm. 
    
\subsubsection{Trial 3: Mean Survival Time Comparison}
The \textit{survival function} can be used to describe the survival probability of a patient at different time. 
The survival functions for GDT and MDT devices can be defined as follow:
    $$ S(t)_{GDT} = \mathbb{P}(T_{GDT}>t)$$
    $$ S(t)_{MDT} = \mathbb{P}(T_{MDT}>t)$$
in which $T_{GDT},T_{MDT}$ are the random variables representing the survival time of a patient in the GDT group and the MDT group.
The survival function can be estimated using the Kaplan-Meier Method \cite{KM_method}. 
From the survival function, the Mean Survival Time $MST_{G},MST_{M}$ can be calculated as:
    $$ MST_{G} = \int_0^{T}S(t)_{GDT}dt$$
    $$ MST_{M} = \int_0^{T}S(t)_{MDT}dt$$
The comparison of the mean survival time is clinically-relevant, but in an actual clinical trial, the comparison does not have statistical significance.

In this trial, we aim to provide statistical significance to the comparison of mean survival time.
The protocol of this pre-clinical trial is different from the previous two trials. 
$n$ virtual patients are created for each simulation instead of one patient in other trials.
The virtual patients are duplicated and are assigned to the GDT and MDT groups.
The following hypothesis compares the mean survival time of the GDT and MDT groups.
    $$H_0:\mathbb{P}(\diamond t\leq T\ MST_{G}>MST_{M}) \geq \mathbb{P}(\diamond t\leq T\  MST_{G}<MST_{M})$$
    $$H_1:\mathbb{P}(\diamond t\leq T\ MST_{G}>MST_{M}) < \mathbb{P}(\diamond t\leq T\  MST_{G}<MST_{M})$$
$2n$ patients are simulated in each iteration and their mean survival times are calculated.
Effectively, each iteration is a RIGHT trial on its own.
The comparison adjudication module is programmed such that:     
    \[ check() =   
    \begin{cases}     
        1       & \quad \text{if } MST_{G}>MST_{M}\\     0  & \quad \text{if } MST_{G}=MST_{M}\\     
        2  & \quad \text{if } MST_{G}<MST_{M}   
    \end{cases} \]   
In this trial we set $n=25$, and after 582 iterations, $H0$ was rejected and $H1$ was accepted with 95\% confidence and 95\% power, suggesting that the new algorithm from GDT has shorter mean event-free survival time compared to the MDT algorithm.
Effectively we have conducted 582 trials with sample size 25, which is infeasible in actual clinical trials.

\subsubsection{Trial 4: Hazard Ratio Comparison}
The \textit{Hazard Ratio} (HR) of the two groups is define as the ratio of two hazard functions $h_{GDT}(t),h_{MDT}(t)$. 
    $$h(t)_{GDT} = \frac{\mathbb{P}(T_{GDT} = t)}{S(t)_{GDT}}, h(t)_{MDT} = \frac{\mathbb{P}(T_{MDT} = t)}{S(t)_{MDT}},HR = \frac{h(t)_{GDT}}{h(t)_{MDT}}$$
The HR reflects the comparative risks for inappropriate therapies between two devices, which can be estimated using the Cox Proportional Hazards Model \cite{Cox}. 

In RIGHT, the hazard ratio was calculated based on the trial result, but its validity cannot be claimed with statistical significance.
In this pre-clinical trial, we aim to compare whether GDT has higher hazard compared to MDT with statistical significance using the following hypothesis:
    $$H_0:\mathbb{P}(\diamond t\leq T\ HR < 1) \geq \mathbb{P}(\diamond t\leq T\  HR > 1)$$
    $$H_1:\mathbb{P}(\diamond t\leq T\ HR < 1) < \mathbb{P}(\diamond t\leq T\  HR > 1)$$
    The comparison adjudication module is programmed such that:     \[ check() =   \begin{cases}     1       & \quad \text{if } HR<1\\     0  & \quad \text{if } HR=1\\     2  & \quad \text{if } HR > 1   \end{cases} \] 
    
The result shows that $H0$ was rejected and $H1$ was accepted with 95\% confidence and 95\% power, suggesting that the new algorithm from GDT has higher hazard compared to the MDT algorithm.
    
    

    

\section{Summary and Future Work}
Clinical trials for medical devices are costly, and many sacrifices have to be made for the trial to be feasible.
Simulations of "virtual patients" cannot replace clinical trials, but can provide useful information that can guide clinical decisions and help the design of clinical trials.
In this paper, we proposed a model-based pre-clinical trial framework using statistical model checking.
Physiological models can provide ground truth for automated adjudication.
Statistical model checker was used to sample physiological parameters, and test clinically-relevant hypothesises with guaranteed statistical significance.
With the proposed framework, we can test hypothesises that are not feasible in clinical trials, or provide statistical significance to the the analysis results during clinical trials.

The next step for this project is to formally prove and quantify the statistical advantage of this framework, and develop templates that can be easily used by researchers in other domains.

\bibliographystyle{splncs04}
\bibliography{bibliography}

\newpage
\appendix

\section{VT-SVT Discrimination Algorithms}
\subsection{RhythmID from Guidant (GDT)}
\begin{figure}      
\centering      
\includegraphics[width=0.3\textwidth]{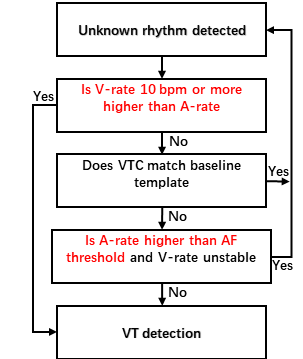}      
\caption{The control flow of the RhythmID algorithm}  
\end{figure}

\begin{figure}      
\centering      
\includegraphics[width=0.6\textwidth]{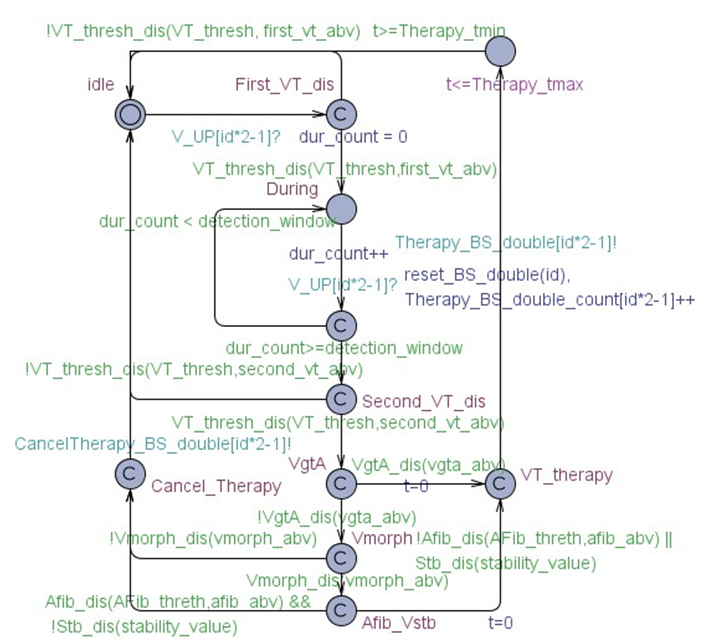}      
\caption{RhythmID algorithm in UPPAAL SMC}  
\end{figure}
\newpage
\subsection{PRLogic from Medtronics (MDT)}
\begin{figure}      
\centering      
\includegraphics[width=0.5\textwidth]{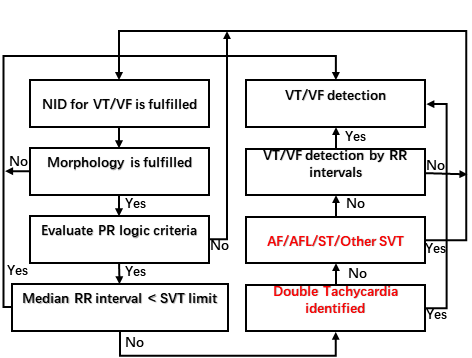}      
\caption{The control flow of the PRLogic algorithm}  
\end{figure}
\begin{figure}

\centering      
\includegraphics[width=1\textwidth]{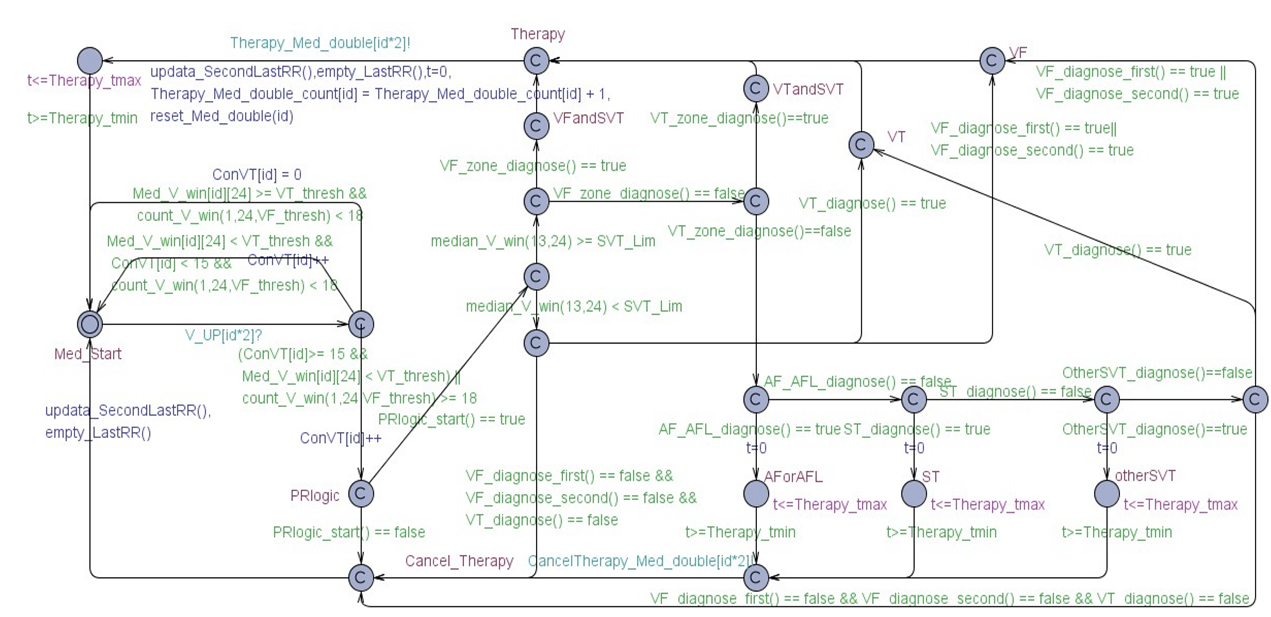}      
\caption{PRLogic algorithm in UPPAAL SMC}  
\end{figure}

\end{document}